# Observer Design for a Class of Systems Described by Differential-Algebraic Equations and Parameter Identification of an Unmeasured Disturbance

**Authors:**

Olga Vladimirovna Oskina, postgraduate student, ov_oskina@itmo.ru, ITMO University, Saint Petersburg, 197101, Russian Federation

Alexey Alekseevich Bobtsov, D.Sc., Professor, bobtsov@itmo.ru, ITMO University, Saint Petersburg, 197101, Russian Federation

**Abstract:** This paper addresses the problem of observer design for a class of linear descriptor systems affected by a certain class of unknown unmatched disturbances. The objective is to estimate the components of the state vector, as well as the unknown parameters of the unmeasured disturbance. To solve this problem, structural assumptions are introduced under which an observer for the dynamic part of the state vector is constructed. Then, based on the obtained state estimate, the disturbance signal is reconstructed, and its unknown parameters are identified. A new parameterization method is proposed for a class of disturbance input signals that depend nonlinearly on unknown parameters, making it possible to obtain a linear regression in the corresponding unknowns. Numerical simulations are presented to demonstrate the effectiveness of the proposed procedures.

**Keywords:** differential-algebraic equations, descriptor systems, adaptive observer, DREM, gradient method, parameter identification

**Introduction.** Differential-algebraic equations are a natural mathematical framework for describing systems in which, unlike systems of ordinary differential equations, it is necessary to account not only for dynamic processes but also for natural relationships among system variables, which are expressed as algebraic equations. In the literature, systems described by differential-algebraic equations are called descriptor systems. The theoretical basis for analyzing the considered class of systems is described in detail in [1-4]. A wide range of applications of such systems, including electrical circuits, mechanical systems and mechatronic devices, and chemical engineering processes, is presented in [5, 6].

The development of theoretically sound and practically applicable observer design methods for dynamic systems described by differential-algebraic equations remains an important research problem with significant practical relevance. In many practical problems, only part of the state variables is measured directly, whereas the remaining variables must be reconstructed from the available system output. The presence of external disturbances or parametric uncertainties introduces additional complexity. Therefore, observer design requires preliminary analysis of the algebraic component of the system, and the direct application of standard methods used for systems of ordinary differential equations is generally not possible. In [7], observers for linear time-varying descriptor systems are considered, and the fundamental importance of consistency between the observer and the structure of the descriptor model is shown. In [8], an observer is proposed that does not require reduction of the system to canonical form and guarantees finite-time convergence of the observation error to zero. The authors of [9,10] consider the problem of observer design for nonlinear descriptor systems using linear matrix inequalities and systematize different types of nonlinearities affecting the observer algorithm design. In [11], a GPEBO-based approach to adaptive observer design for linear time-varying descriptor systems is proposed. The authors of [11], as well as those of [12], consider systems subject to disturbances and note that many state estimation methods rely on a reduction to ordinary differential equations, which leads to a loss of full system information and affects the estimation accuracy. A related direction is the stabilization of a system in the presence of unmeasured matched and unmatched external disturbances. In [13, 14], the internal model principle is used to compensate external disturbances, where the descriptor formalism is employed as a convenient mathematical framework for the case in which not the entire output but only part of it is regulated, giving rise to a transfer function whose numerator degree exceeds that of the denominator.

Thus, an analysis of the available works shows that the observer design problem for systems having the structure of a system of differential-algebraic equations in the presence of disturbances is relevant. Parametric identification of the unknown disturbance parameters can be used in the subsequent problem of compensating these disturbances and constructing a control algorithm.

**Problem Statement.** Consider a linear time-invariant model described by the following system of differential-algebraic equations:

$$\begin{gathered}\boldsymbol{E}\dot{\boldsymbol{x}}(t) = \boldsymbol{A}\boldsymbol{x}(t) + \boldsymbol{B}_{\boldsymbol{u}}\boldsymbol{u}(t) + \boldsymbol{B}_{\delta}\delta(t),\ \boldsymbol{x}(0) = \boldsymbol{x}_0 \in \mathbb{R}^n, t \geq 0,\\ \boldsymbol{y}(t) = \boldsymbol{C}\boldsymbol{x}(t),\end{gathered} \tag{1}$$

where $\boldsymbol{x} = (\boldsymbol{x}_a \quad \boldsymbol{x}_b)^{\mathrm{T}}$ is the unmeasured state vector; $\boldsymbol{x}_a \in \mathbb{R}^{n_a}$ is the dynamic component of the state vector; $\boldsymbol{x}_b \in \mathbb{R}^{n_b}$ is the algebraic component of the state vector, $n = n_a + n_b$, $\boldsymbol{u}(t) \in \mathbb{R}$ is the known input signal, $\boldsymbol{y}(t) \in \mathbb{R}$ is the measured output signal, $\delta(t) \in \mathbb{R}$ is an unknown disturbance input, matrices $\boldsymbol{E} \in \mathbb{R}^{n\times n}$, $\boldsymbol{A} \in \mathbb{R}^{n\times n}$, $\boldsymbol{B}_{\boldsymbol{u}} \in \mathbb{R}^{n\times 1}$, $\boldsymbol{B}_{\delta} \in \mathbb{R}^{n\times 1}$ and $\boldsymbol{C} \in \mathbb{R}^{1\times n}$ are known and constant.

The following assumptions are introduced within the stated problem.

*Assumption 1.* The matrix $\boldsymbol{E} \in \mathbb{R}^{n\times n}$ is singular and has the following form:

$$\boldsymbol{E} = \begin{bmatrix} \boldsymbol{I}_{n_a} & 0_{n_a\times n_b} \\ 0_{n_b\times n_a} & 0_{n_b\times n_b} \end{bmatrix}$$

In view of this assumption, system (1) can be rewritten as [11]:

$$\begin{cases} \dot{\boldsymbol{x}}_a(t) = \boldsymbol{A}_{11}\boldsymbol{x}_a(t) + \boldsymbol{A}_{12}\boldsymbol{x}_b(t) + \boldsymbol{B}_{11}\boldsymbol{u}(t) + \boldsymbol{B}_{12}\delta(t), \\ 0 = \boldsymbol{A}_{21}\boldsymbol{x}_a(t) + \boldsymbol{A}_{22}\boldsymbol{x}_b(t) + \boldsymbol{B}_{21}\boldsymbol{u}(t) + \boldsymbol{B}_{22}\delta(t), \\ \boldsymbol{y} = \boldsymbol{C}_1\boldsymbol{x}_a(t) + \boldsymbol{C}_2\boldsymbol{x}_b(t). \end{cases} \tag{2}$$

where $\boldsymbol{A} = \begin{bmatrix} \boldsymbol{A}_{11} & \boldsymbol{A}_{12} \\ \boldsymbol{A}_{21} & \boldsymbol{A}_{22} \end{bmatrix}$, $\quad \boldsymbol{B}_{\boldsymbol{u}} = \begin{bmatrix} \boldsymbol{B}_{11} \\ \boldsymbol{B}_{21} \end{bmatrix}$, $\quad \boldsymbol{B}_{\delta} = \begin{bmatrix} \boldsymbol{B}_{12} \\ \boldsymbol{B}_{22} \end{bmatrix}$, $\quad \boldsymbol{C} = [\boldsymbol{C}_1 \quad \boldsymbol{C}_2]$

*Assumption 2.* The algebraic equation must be solvable with respect to $\boldsymbol{x}_b$:

$$\operatorname{rank} \boldsymbol{A}_{22} = n_b \tag{3}$$

*Assumption 3.* The disturbance input $\delta(t)$ can be represented as follows [15]:

$$\delta(t) = \sum_{i=1}^{r} \alpha_i\,(t) f_i(\theta_i, t) + \delta_0(t) \tag{4}$$

where $\delta_0(t)$ is an unknown but bounded irregular component, $\alpha_i(t)$ is a known function, $\theta_i$ is an unknown constant parameter, and the function $f_i(\theta_i, t)$ can be parameterized using positive delays $D_j$ $(D_1 \neq D_2 \neq \cdots \neq D_S)$ of known values in the following form:

$$f_i(\theta_i, t - D_j) = \sum_{k=1}^{v_i} K_{ijk}(\theta_i)\beta_{ik}(\theta_i, t), \qquad j = 1, \dots, s \tag{5}$$

where $K_{ijk}(\theta_i)$ are coefficients depending on the unknown parameters, with the inverse transformation $\theta_i = f^{-1}(K_{ijk})$ assumed to exist, $\beta_{ik}(\theta_i, t)$ are basis functions used to describe the signal. The total number of these functions is equal to $\ell = \sum_{i=1}^{r} v_i$. The index $i$ enumerates the disturbance component, the index $k$ enumerates the basis function within this component, and the index $j$ enumerates the delay. Moreover, $s \geq \ell + 1$.

The component $\boldsymbol{x}_b$ is computable. Therefore, the main objective is to construct an observer for $\widehat{\boldsymbol{x}}_a$, such that

$$\lim_{t\to\infty}(\widehat{\boldsymbol{x}}_a(t) - \boldsymbol{x}_a(t)) = 0. \tag{6}$$

After the state vector estimate is obtained, the unknown disturbance parameters can be recovered.

**Main Result.** According to Assumption 2, we obtain the following condition

$$\det(\boldsymbol{A}_{22}^T \boldsymbol{A}_{22}) \neq 0 \tag{7}$$

Now, from the algebraic equation of system (2), we can express $\boldsymbol{x}_b$ as follows:

$$\boldsymbol{x}_b(t) = -\boldsymbol{A}_{22}^{-1}(\boldsymbol{A}_{21}\boldsymbol{x}_a(t) + \boldsymbol{B}_{21}\boldsymbol{u}(t) + \boldsymbol{B}_{22}\delta(t)) \tag{8}$$

where $\boldsymbol{A}_{22}^{-1} = (\boldsymbol{A}_{22}^T \boldsymbol{A}_{22})^{-1} \boldsymbol{A}_{22}^T$

Substitution of expression (8) into the first equation of system (2) gives a differential equation with respect to $\boldsymbol{x}_a$

$$\dot{\boldsymbol{x}}_a(t) = \boldsymbol{A}_a \boldsymbol{x}_a(t) + \boldsymbol{B}_{1a}\boldsymbol{u}(t) + \boldsymbol{B}_{2a}\delta(t) \tag{9}$$

where

$$\begin{aligned} \boldsymbol{A}_a &= \boldsymbol{A}_{11} - \boldsymbol{A}_{12}\boldsymbol{A}_{22}^{-1}\boldsymbol{A}_{21}, \\ \boldsymbol{B}_{1a} &= \boldsymbol{B}_{11} - \boldsymbol{A}_{12}\boldsymbol{A}_{22}^{-1}\boldsymbol{B}_{21}, \\ \boldsymbol{B}_{2a} &= \boldsymbol{B}_{12} - \boldsymbol{A}_{12}\boldsymbol{A}_{22}^{-1}\boldsymbol{B}_{22}. \end{aligned}$$

Thus, the descriptor system has been reduced to an ordinary differential equation with respect to the state $\boldsymbol{x}_a$.

Now, substituting expression (8) into the output equation, we obtain:

$$\boldsymbol{y}(t) = \boldsymbol{C}_3 \boldsymbol{x}_a(t) + \boldsymbol{C}_4 \boldsymbol{u}(t) + \boldsymbol{C}_5 \delta(t) \tag{10}$$

where

$$\boldsymbol{C}_3 = \boldsymbol{C}_1 - \boldsymbol{C}_2\boldsymbol{A}_{22}^{-1}\boldsymbol{A}_{21},$$
$$\boldsymbol{C}_4 = -\boldsymbol{C}_2\boldsymbol{A}_{22}^{-1}\boldsymbol{B}_{21},$$
$$\boldsymbol{C}_5 = -\boldsymbol{C}_2\boldsymbol{A}_{22}^{-1}\boldsymbol{B}_{22}.$$

*Assumption 4.* The reduced output equation has no direct disturbance channel, so:

$$\boldsymbol{C}_2\boldsymbol{A}_{22}^{-1}\boldsymbol{B}_{22} = 0 \tag{11}$$

Under this assumption, we obtain that $\boldsymbol{C}_5 = 0$, and therefore the output equation has the form:

$$\boldsymbol{y}(t) = \boldsymbol{C}_3\boldsymbol{x}_a(t) + \boldsymbol{C}_4\boldsymbol{u}(t) \tag{12}$$

Provided that the matrices $\boldsymbol{C}_3$ and $\boldsymbol{C}_4$ are constant and the functions $\boldsymbol{y}(t)$ and $\boldsymbol{u}(t)$ are smooth, differentiating the output equation yields:

$$\dot{\boldsymbol{y}}(\mathrm{t}) = \boldsymbol{C}_3\boldsymbol{A}_a\boldsymbol{x}_a(t) + \boldsymbol{C}_3\boldsymbol{B}_{1a}\boldsymbol{u}(t) + \boldsymbol{C}_3\boldsymbol{B}_{2a}\delta(t) + \boldsymbol{C}_4\dot{\boldsymbol{u}}(t) \tag{13}$$

*Assumption 5.* The following relation holds for the matrices $\boldsymbol{C}_3$ and $\boldsymbol{B}_{2a}$:

$$\boldsymbol{C}_3\boldsymbol{B}_{2a} \neq 0 \tag{14}$$

Then the disturbance input can be expressed from (13) as follows:

$$\delta(t) = \frac{1}{\boldsymbol{C}_3\boldsymbol{B}_{2a}}(\dot{\boldsymbol{y}}(t) - \boldsymbol{C}_3\boldsymbol{A}_a\boldsymbol{x}_a(t) - \boldsymbol{C}_3\boldsymbol{B}_{1a}\boldsymbol{u}(t) - \boldsymbol{C}_4\dot{\boldsymbol{u}}(t)) \tag{15}$$

Now, substituting the expression for the disturbance into (9), we obtain:

$$\dot{\boldsymbol{x}}_a(t) = \boldsymbol{M}_a\boldsymbol{x}_a(t) + \boldsymbol{N}\boldsymbol{u}(t) + \boldsymbol{F}(\dot{\boldsymbol{y}}(\mathrm{t}) - \boldsymbol{C}_4\dot{\boldsymbol{u}}(t)) \tag{16}$$

where

$$\boldsymbol{M}_a = \left(\boldsymbol{I} - \frac{\boldsymbol{B}_{2a}\boldsymbol{C}_3}{\boldsymbol{C}_3\boldsymbol{B}_{2a}}\right)\boldsymbol{A}_a,$$
$$\boldsymbol{N} = \left(\boldsymbol{I} - \frac{\boldsymbol{B}_{2a}\boldsymbol{C}_3}{\boldsymbol{C}_3\boldsymbol{B}_{2a}}\right)\boldsymbol{B}_{1a},$$
$$\boldsymbol{F} = \frac{\boldsymbol{B}_{2a}}{\boldsymbol{C}_3\boldsymbol{B}_{2a}}.$$

The introduced structural assumptions yield a differential equation for the component $\boldsymbol{x}_a$ that does not contain the disturbance signal.

To obtain the estimate $\widehat{\boldsymbol{x}}_a$, the following unknown-input observer structure is used [16]:

$$\dot{\widehat{\boldsymbol{x}}}_a(t) = \boldsymbol{M}_a\widehat{\boldsymbol{x}}_a(t) + \boldsymbol{N}\boldsymbol{u}(t) + \boldsymbol{F}(\dot{\boldsymbol{y}}(t) - \boldsymbol{C}_4\dot{\boldsymbol{u}}(t)) + \boldsymbol{L}(\boldsymbol{y}(t) - \boldsymbol{C}_3\widehat{\boldsymbol{x}}_a(t) - \boldsymbol{C}_4\boldsymbol{u}(t)) \tag{17}$$

where $(\boldsymbol{M}_a, \boldsymbol{C}_3)$ is detectable, and, to satisfy the target condition (6), $\boldsymbol{L}$ is chosen so that $\boldsymbol{M}_a - \boldsymbol{L}\boldsymbol{C}_3$ is Hurwitz.

The observer contains derivatives of the output signal $\dot{\boldsymbol{y}}$ and of the control input $\dot{\boldsymbol{u}}$. To avoid differentiation, one can perform a change of coordinates and introduce the variable

$$\boldsymbol{z}(t) = \hat{\boldsymbol{x}}_a(t) - \boldsymbol{F}(\boldsymbol{y}(t) - \boldsymbol{C}_4\boldsymbol{u}(t))$$

Then observer (17) takes the following implementable form:

$$\begin{cases} \dot{\boldsymbol{z}}(t) = (\boldsymbol{M}_a - \boldsymbol{L}\boldsymbol{C}_3)\hat{\boldsymbol{x}}_a(t) + \boldsymbol{N}\boldsymbol{u}(t) + \boldsymbol{L}(\boldsymbol{y}(t) - \boldsymbol{C}_4\boldsymbol{u}(t)), \\ \hat{\boldsymbol{x}}_a(t) = \boldsymbol{z}(t) + \boldsymbol{F}(\boldsymbol{y}(t) - \boldsymbol{C}_4\boldsymbol{u}(t)). \end{cases} \tag{18}$$

After obtaining the estimate of the state-vector component $\hat{\boldsymbol{x}}_a$, one can reconstruct the disturbance input signal $\hat{\delta}$ by substituting the estimate $\hat{\boldsymbol{x}}_a$ into expression (15). Since the derivatives $\dot{\boldsymbol{y}}(t)$ and $\dot{\boldsymbol{u}}(t)$ are not measured directly, a stable filter $D_\lambda(p) = \frac{\lambda_f p}{p+\lambda_f}, \lambda_f > 0$ is implemented to recover them. Then we obtain:

$$\begin{aligned} \dot{q}_{\boldsymbol{y}} &= -\lambda_f q_{\boldsymbol{y}} + \lambda_f \boldsymbol{y}, \qquad & \hat{\boldsymbol{y}}_{\lambda_f} &= \lambda_f \boldsymbol{y} - \lambda_f q_{\boldsymbol{y}}, \\ \dot{q}_{\boldsymbol{u}} &= -\lambda_f q_{\boldsymbol{u}} + \lambda_f \boldsymbol{u}, \qquad & \hat{\boldsymbol{u}}_{\lambda_f} &= \lambda_f \boldsymbol{u} - \lambda_f q_{\boldsymbol{u}}, \end{aligned} \tag{19}$$

Thus, the final expression for obtaining the disturbance estimate has the form:

$$\hat{\delta} = \frac{1}{\boldsymbol{C}_3\boldsymbol{B}_{2a}}\left(\hat{\boldsymbol{y}}_{\lambda_f} - \boldsymbol{C}_3\boldsymbol{A}_a\hat{\boldsymbol{x}}_a - \boldsymbol{C}_3\boldsymbol{B}_{1a}\boldsymbol{u} - \boldsymbol{C}_4\hat{\boldsymbol{u}}_{\lambda_f}\right) \tag{20}$$

After obtaining $\hat{\boldsymbol{x}}_a$ and reconstructing the disturbance signal, one can identify the unknown parameters $\theta_i$.

To demonstrate the proposed procedure for estimating the unknown parameters, the irregular component $\delta_o(t)$ may be omitted. The procedure to be used when this component is present is described in [15].

In the derivation below, the reconstructed signal is denoted by $\delta(t)$ to keep the expressions uncluttered.

For each delay, the general form of the disturbance (4) can be written according to the expression (5):

$$\delta(t-D_j)=\sum_{i=1}^{r}\sum_{k=1}^{v_i}\alpha_i(t-D_j)K_{ijk}(\theta_i)\beta_{ik}(\theta_i,t)=\varphi_j^T(\theta,t)\boldsymbol{Q}(\theta,t) \tag{21}$$

where $\boldsymbol{Q}(\theta,t)=\begin{bmatrix}\beta_{11}(\theta_1,t)\\ \vdots \\ \beta_{1v_1}(\theta_1,t)\\ \vdots \\ \beta_{r1}(\theta_r,t)\\ \vdots \\ \beta_{rv_r}(\theta_r,t)\end{bmatrix}$ is the vector of basis functions, $\boldsymbol{Q}(\theta,t)\in\mathbb{R}^{\ell}$,

$\varphi_j(\theta,t)=\begin{bmatrix}\alpha_1(t-D_j)K_{1j1}(\theta_1)\\ \vdots \\ \alpha_1(t-D_j)K_{1jv_1}(\theta_1)\\ \vdots \\ \alpha_r(t-D_j)K_{rj1}(\theta_r)\\ \vdots \\ \alpha_r(t-D_j)K_{rjv_r}(\theta_r)\end{bmatrix}$ is the vector containing the known time-varying factors $\alpha_i(t-D_j)$ and the unknown coefficients $K_{ijk}(\theta_i)$.

To construct a linear regression, we propose a generalization of the DREM procedure [17], since in the present problem formulation the unknown parameters enter the regression equation nonlinearly. At the extension stage, $\ell$ number of delays $D_1,\dots,D_\ell$ are introduced, which yields the extended regression:

$$\boldsymbol{\Upsilon}(t)=\boldsymbol{\Phi}(\theta,t)\boldsymbol{Q}(\theta,t) \tag{22}$$

where $\boldsymbol{\Upsilon}(t)=\begin{bmatrix}\delta(t-D_1)\\ \vdots \\ \delta(t-D_\ell)\end{bmatrix}$, $\boldsymbol{\Phi}(\theta,t)=\begin{bmatrix}\varphi_1^T(\theta,t)\\ \vdots \\ \varphi_\ell^T(\theta,t)\end{bmatrix}$.

We perform the mixing procedure, which gives:

$$\Delta(\theta,t)\boldsymbol{Q}(\theta,t)=\mathrm{adj}\{\boldsymbol{\Phi}(\theta,t)\}\boldsymbol{\Upsilon}(t) \tag{23}$$

where $\Delta(\theta,t)=\det\boldsymbol{\Phi}(\theta,t)$

We introduce one more delay $D_{\ell+1}$, perform a series of algebraic transformations, and obtain:

$$\Delta(\theta,t)\delta(t-D_{\ell+1})=\varphi_{\ell+1}^T(\theta,t)\mathrm{adj}\{\boldsymbol{\Phi}(\theta,t)\}\boldsymbol{\Upsilon}(t) \tag{24}$$

where $\Delta(\theta,t)=\det\boldsymbol{\Phi}(\theta,t)$, $\delta(t-D_{\ell+1})=\varphi_{\ell+1}^T(\theta,t)\boldsymbol{Q}(\theta,t)$

A parameterization is performed to extract the unknown constant parameters $K_{ijk}(\theta_i)$. To this end, we substitute into (24) the expressions for $\Delta$, $\varphi_{\ell+1}$, $\mathrm{adj}\{\boldsymbol{\Phi}\}$ and $\boldsymbol{\Upsilon}$ and obtain that the left- and right-hand sides of (24) give a finite sum of products of two types: combinations $\mu_c(t)$ of known signals $\alpha_i(t-D_j)$ and $\delta(t-D_j)$ and combinations $\chi_c(\theta)$ of unknown coefficients $K_{ijk}(\theta_i)$, $c = 0, \dots, q$. Therefore, expression (24) can be written in the following form:

$$\mu_0(t)\chi_0(\theta) + \mu_1(t)\chi_1(\theta) + \cdots + \mu_q(t)\chi_q(\theta) = 0 \tag{25}$$

Next, we choose a nonzero combination $\chi(\theta)$ and divide expression (25) by it. We obtain a linear regression equation:

$$p = m^T\boldsymbol{\eta} \tag{26}$$

where $p(t) = -\mu_0(t)$, $m(t) = \begin{bmatrix} \mu_1(t) \\ \vdots \\ \mu_q(t) \end{bmatrix}$, $\boldsymbol{\eta} = \begin{bmatrix} \eta_1 \\ \vdots \\ \eta_q \end{bmatrix}$, $\eta_c = \frac{\chi_c(\theta)}{\chi_0(\theta)}$

Once the linear regression equation is obtained, the vector of unknown parameters $\boldsymbol{\eta}$ can be estimated using any standard method, such as a gradient algorithm or the least-squares method. The parameters $\theta_i$ can be recovered from the corresponding algebraic relations between $\boldsymbol{\eta}$ and $\theta_i$.

Thus, the proposed parameterization procedure enables the original disturbance model of the considered class, nonlinear in the unknown parameters, to be transformed into a linear regression equation with respect to new constant combinations of these parameters.

To illustrate the proposed methods, consider the following academic example.

**Example.** Let system (1) have the form:

$$\begin{cases} \boldsymbol{E}\dot{\boldsymbol{x}}(t) = \boldsymbol{A}\boldsymbol{x}(t) + \boldsymbol{B_u}\boldsymbol{u}(t) + \boldsymbol{B}_\delta\delta(t), \\ \boldsymbol{y}(t) = \boldsymbol{C}\boldsymbol{x}(t), \end{cases}$$

where $\boldsymbol{x} \in \mathbb{R}^3$; $\boldsymbol{x} = \begin{bmatrix} \boldsymbol{x}_a \\ \boldsymbol{x}_b \end{bmatrix}$, $\boldsymbol{x}_a = \begin{bmatrix} x_1 \\ x_2 \end{bmatrix}$, $\boldsymbol{E} = \begin{bmatrix} \boldsymbol{I}_{n_a} & 0_{n_a \times n_b} \\ 0_{n_b \times n_a} & 0_{n_b \times n_b} \end{bmatrix} = \begin{bmatrix} 1 & 0 & 0 \\ 0 & 1 & 0 \\ 0 & 0 & 0 \end{bmatrix}$,

$\boldsymbol{A} = \begin{bmatrix} \boldsymbol{A}_{11} & \boldsymbol{A}_{12} \\ \boldsymbol{A}_{21} & \boldsymbol{A}_{22} \end{bmatrix} = \begin{bmatrix} 0 & 1 & 1 \\ 0 & -2 & 1 \\ 1 & 0 & 1 \end{bmatrix}$, $\boldsymbol{B_u} = \begin{bmatrix} \boldsymbol{B}_{11} \\ \boldsymbol{B}_{21} \end{bmatrix} = \begin{bmatrix} 0 \\ 2 \\ 1 \end{bmatrix}$, $\boldsymbol{B}_\delta = \begin{bmatrix} \boldsymbol{B}_{12} \\ \boldsymbol{B}_{22} \end{bmatrix} = \begin{bmatrix} 1 \\ 0 \\ 0 \end{bmatrix}$,

$\boldsymbol{C} = [\boldsymbol{C}_1 \quad \boldsymbol{C}_2] = [2 \quad 0 \quad 1], \boldsymbol{u}(t) = \sin(t)$

In form (2), the system is:

$$\begin{cases} \dot{\boldsymbol{x}}_1 = \boldsymbol{x}_2 + \boldsymbol{x}_b + \delta \\ \dot{\boldsymbol{x}}_2 = -2\boldsymbol{x}_2 + \boldsymbol{x}_b + 2\boldsymbol{u} \\ 0 = \boldsymbol{x}_1 + \boldsymbol{x}_b + \boldsymbol{u} \\ \boldsymbol{y} = 2\boldsymbol{x}_1 + \boldsymbol{x}_b \end{cases}, \qquad \boldsymbol{x}_b(0) = -\boldsymbol{x}_1(0) - \boldsymbol{u}(0) \tag{27}$$

The reduced model has the form:

$$\begin{gathered} \dot{\boldsymbol{x}}_\mathrm{a} = \boldsymbol{A}_a\boldsymbol{x}_a + \boldsymbol{B}_{1a}\boldsymbol{u} + \boldsymbol{B}_{2a}\delta \\ \boldsymbol{y} = \boldsymbol{C}_3\boldsymbol{x}_a + \boldsymbol{C}_4\boldsymbol{u} \end{gathered} \tag{28}$$

where $\boldsymbol{A}_a = \begin{bmatrix} -1 & 1 \\ -1 & -2 \end{bmatrix}, \boldsymbol{B}_{1a} = \begin{bmatrix} -1 \\ 1 \end{bmatrix}, \boldsymbol{B}_{2a} = \begin{bmatrix} 1 \\ 0 \end{bmatrix}, \boldsymbol{C}_3 = [1 \quad 0], \boldsymbol{C}_4 = -1$

The observer model $\hat{\boldsymbol{x}}_a$ has the form:

$$\begin{cases} \dot{\boldsymbol{z}} = (\boldsymbol{M}_a - \boldsymbol{L}\boldsymbol{C}_3)\hat{\boldsymbol{x}}_a + \boldsymbol{N}\boldsymbol{u} + \boldsymbol{L}(\boldsymbol{y} - \boldsymbol{C}_4\boldsymbol{u}) \\ \hat{\boldsymbol{x}}_a = \boldsymbol{z} + \boldsymbol{F}(\boldsymbol{y} - \boldsymbol{C}_4\boldsymbol{u}) \end{cases} \tag{29}$$

where $\boldsymbol{F} = \begin{bmatrix} 1 \\ 0 \end{bmatrix}$, $\boldsymbol{M}_a = \begin{bmatrix} 0 & 0 \\ -1 & -2 \end{bmatrix}, \boldsymbol{N} = \begin{bmatrix} 0 \\ 1 \end{bmatrix}$

After obtaining the estimate $\hat{\boldsymbol{x}}_a$, the disturbance is reconstructed from the output equation:

$$\hat{\delta} = \hat{\dot{\boldsymbol{y}}} + \hat{\boldsymbol{x}}_1 - \hat{\boldsymbol{x}}_2 + \boldsymbol{u} + \hat{\dot{\boldsymbol{u}}} \tag{30}$$

where the derivatives $\hat{\dot{\boldsymbol{u}}}$ and $\hat{\dot{\boldsymbol{y}}}$ are implemented using linear filters (19).

After reconstructing the disturbance signal, one can proceed to estimating its unknown parameters. The disturbance signal is chosen as a sum of two exponential components with known time-varying factors:

$$\delta(t) = \alpha_1(t)e^{\theta_1 t} + \alpha_2(t)e^{\theta_2 t} \tag{31}$$

where $\alpha_1(t) = 1.35 + 0.25\sin(1.7t), \alpha_2(t) = 1.1 + 0.20\cos(2.3t), \theta_1, \theta_2$ are unknown constant parameters.

We perform the extension and mixing operations according to procedure (21)-(24). We obtain the following expression:

$$\Delta\delta(t - D_3) = \varphi_3^T \mathrm{adj}\{\boldsymbol{\Phi}\}\boldsymbol{\Upsilon} \tag{32}$$

where $\Delta = \det\boldsymbol{\Phi}$, $\boldsymbol{\Phi} = \begin{bmatrix} \varphi_1^T \\ \varphi_2^T \end{bmatrix}, \boldsymbol{\Upsilon} = \begin{bmatrix} \delta(t - D_1) \\ \delta(t - D_2) \end{bmatrix}, \delta\left(t - D_j\right) = \varphi_j^T\boldsymbol{Q},$

$\varphi_j^T(\theta, t) = [\alpha_1(t - D_j)K_{ij}(\theta_1) \quad \alpha_2(t - D_j)K_{ij}(\theta_2)]$, $K_{ij}(\theta_i) = \mathrm{e}^{-\theta_i D_j}, i = 1,2$, $j = 1,2,3$

A parameterization is introduced to isolate the unknown constant coefficients depending on the unknown disturbance parameters. Substituting the explicit expressions for the known components into (32), expanding the brackets, and grouping the terms yields a regression equation that is linear with respect to new unknown parameter combinations:

$$p = \boldsymbol{m}^T \boldsymbol{\eta}$$

where $p = -\mu_1, \boldsymbol{m} = [\mu_2 \quad \mu_3 \quad \mu_4 \quad \mu_5 \quad \mu_6], \boldsymbol{\eta} = [\eta_2 \quad \eta_3 \quad \eta_4 \quad \eta_5 \quad \eta_6]$,

$\mu_1 = \alpha_1(t - D_1)\alpha_2(t - D_2)\delta(t - D_3), \mu_2 = -\alpha_2(t - D_1)\alpha_1(t - D_2)\delta(t - D_3)$,

$\mu_3 = -\alpha_1(t - D_3)\alpha_2(t - D_2)\delta(t - D_1), \mu_4 = \alpha_1(t - D_3)\alpha_2(t - D_1)\delta(t - D_2)$,

$\mu_5 = \alpha_2(t - D_3)\alpha_1(t - D_2)\delta(t - D_1)$, $\mu_6 = -\alpha_2(t - D_3)\alpha_1(t - D_1)\delta(t - D_2)$.

$$\eta_2 = \frac{K_{21}(\theta_2)K_{12}(\theta_1)}{K_{11}(\theta_1)K_{22}(\theta_2)}, \eta_3 = \frac{K_{13}(\theta_1)K_{22}(\theta_2)}{K_{11}(\theta_1)K_{22}(\theta_2)}, \eta_4 = \frac{K_{13}(\theta_1)K_{21}(\theta_2)}{K_{11}(\theta_1)K_{22}(\theta_2)},$$

$$\eta_5 = \frac{K_{23}(\theta_2)K_{12}(\theta_1)}{K_{11}(\theta_1)K_{22}(\theta_2)}, \eta_6 = \frac{K_{23}(\theta_2)K_{11}(\theta_1)}{K_{11}(\theta_1)K_{22}(\theta_2)}.$$

We estimate the parameters of the regression model using the recursive least-squares method (see, for example, [18]) in the following form

$$\dot{\hat{\boldsymbol{\eta}}} = \boldsymbol{P}(t)\boldsymbol{m}(t)(p - \boldsymbol{m}(t)^T \hat{\boldsymbol{\eta}}) \tag{33}$$

where the covariance matrix $\boldsymbol{P}(t)$ changes according to the following law

$$\dot{\boldsymbol{P}}(t) = -\boldsymbol{P}(t)\boldsymbol{m}(t)\boldsymbol{m}(t)^T \boldsymbol{P}(t) \tag{34}$$

where $\boldsymbol{P}(0) = \frac{1}{p_0}\boldsymbol{I}, p_0 > 0$

Having obtained the estimate $\boldsymbol{\eta}$, we obtain the estimates of the parameters $\theta_1, \theta_2$, according to the expressions:

$$\hat{\theta}_1 = \frac{\ln(\hat{\eta}_3)}{D_1 - D_3}; \hat{\theta}_2 = \frac{\ln(\hat{\eta}_6)}{D_2 - D_3} \tag{35}$$

Since the regression is formed from the reconstructed signal $\hat{\delta}$ and its time-shifted values, the initial transients of observer (29) and filters (19) should be excluded from the regression equation. Otherwise, the adaptive algorithm (33) will use not only the informative part of the signal but also the reconstruction error in $\hat{\delta}$, which may bias the estimate $\hat{\boldsymbol{\eta}}$. Therefore, identification should be started after the calculated time:

$$T_{id} = T_{\text{obs}} + T_f + D_{max} \tag{36}$$

where $D_{max} = \max_j D_j$ is the maximum delay value used in constructing the extended regression.

The time $T_{\text{obs}}$ is determined by the convergence rate of observer (29). Since the estimation error satisfies $\dot{\tilde{\boldsymbol{x}}}_a = (\boldsymbol{M}_a - \boldsymbol{L}\boldsymbol{C}_3)\tilde{\boldsymbol{x}}_a$, by choosing $\boldsymbol{L}$, we set:

$$T_{\text{obs}} = \frac{\ln(1/\varepsilon_{\text{obs}})}{\lambda_{obs}}, \qquad \lambda_{obs} = \min_i |\text{Re}\{\lambda_i(\boldsymbol{M}_a - \boldsymbol{L}\boldsymbol{C}_3)\}| \tag{37}$$

where $\varepsilon_{\text{obs}} \in (0,1)$ specifies the admissible fraction of the initial estimation error remaining by the time identification is enabled.

When derivatives are implemented through first-order filters, the free component of the filter error decays as $\mathrm{e}^{-\lambda_f t}$. Hence, the filter transient decay time is chosen as:

$$T_f = \frac{\ln(1/\varepsilon_f)}{\lambda_f}, \tag{38}$$

where $\varepsilon_f \in (0,1)$ specifies the admissible fraction of the initial filter transient component remaining by the time identification is enabled.

We perform a computer simulation of the estimation procedure for $\hat{\boldsymbol{x}}_a$, $\hat{\delta}$, $\hat{\theta}_1$ and $\hat{\theta}_2$ with initial conditions $\boldsymbol{x}_a(0) = \begin{bmatrix} 1 \\ -1 \end{bmatrix}, \hat{\boldsymbol{x}}_a(0) = \begin{bmatrix} 0 \\ 0 \end{bmatrix}, \boldsymbol{z}(0) = \begin{bmatrix} -1 \\ 0 \end{bmatrix}, \boldsymbol{x}_b(0) = -1$, with disturbance parameters: $\theta_1 = -0.1, \theta_2 = -0.3$, with matrix $\boldsymbol{L} = \begin{bmatrix} 4 \\ 0 \end{bmatrix}$, filter parameter $\lambda_f = 10^3$, delays $D_1 = 1, D_2 = 2, D_3 = 3$, and initial condition of the covariance matrix $\boldsymbol{P}(0) = 10^3\mathrm{I}$.

The simulation results for the algorithm estimating the unknown component of the state vector $\hat{\boldsymbol{x}}_a$ and the disturbance input $\hat{\delta}$ are shown in Figs. 1 and 2, respectively. The simulation results for the algorithm estimating the parameters of the disturbance $\hat{\delta}$ are shown in Fig. 3.

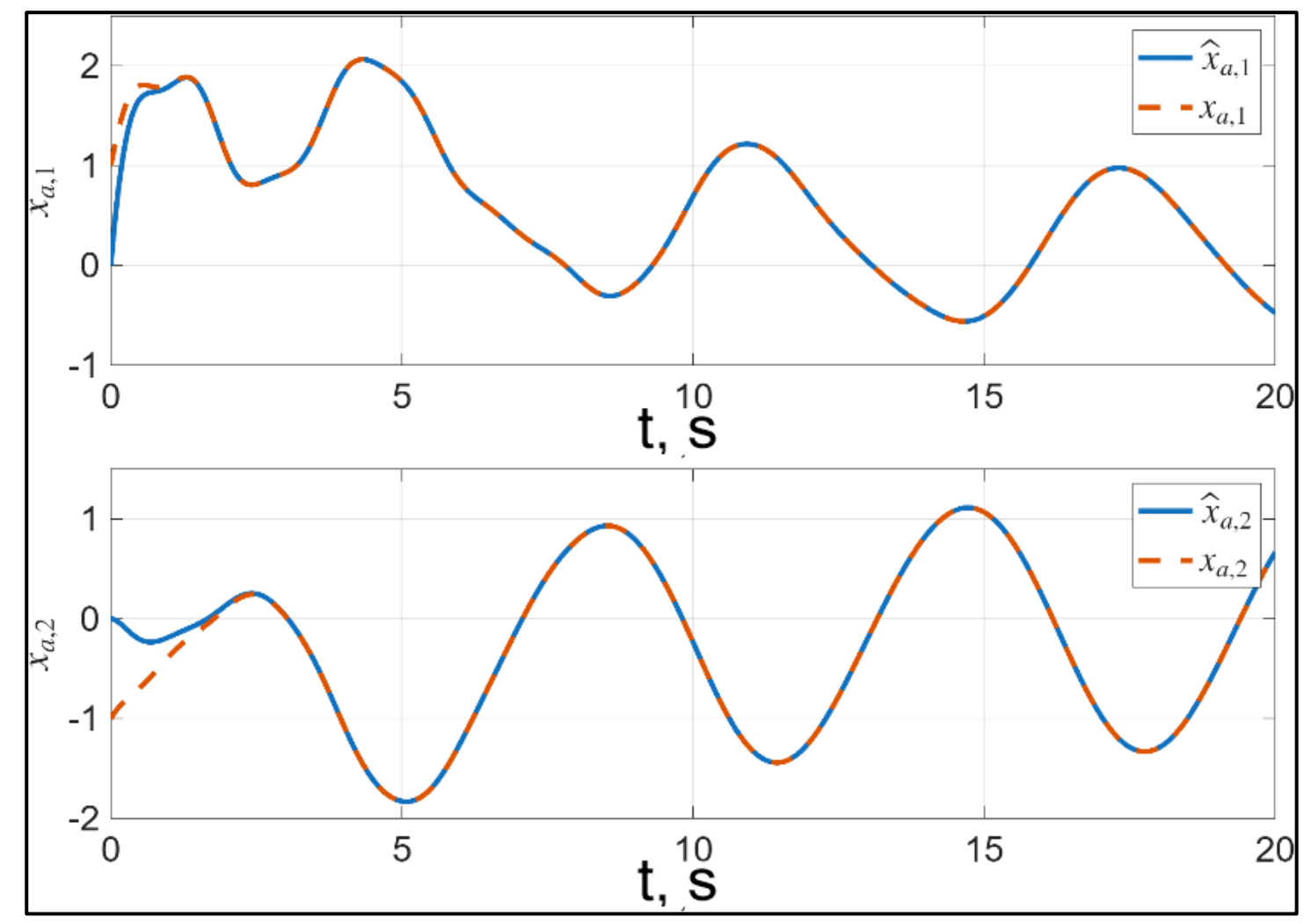

Figure 1 – Estimation results for $\hat{\boldsymbol{x}}_a$

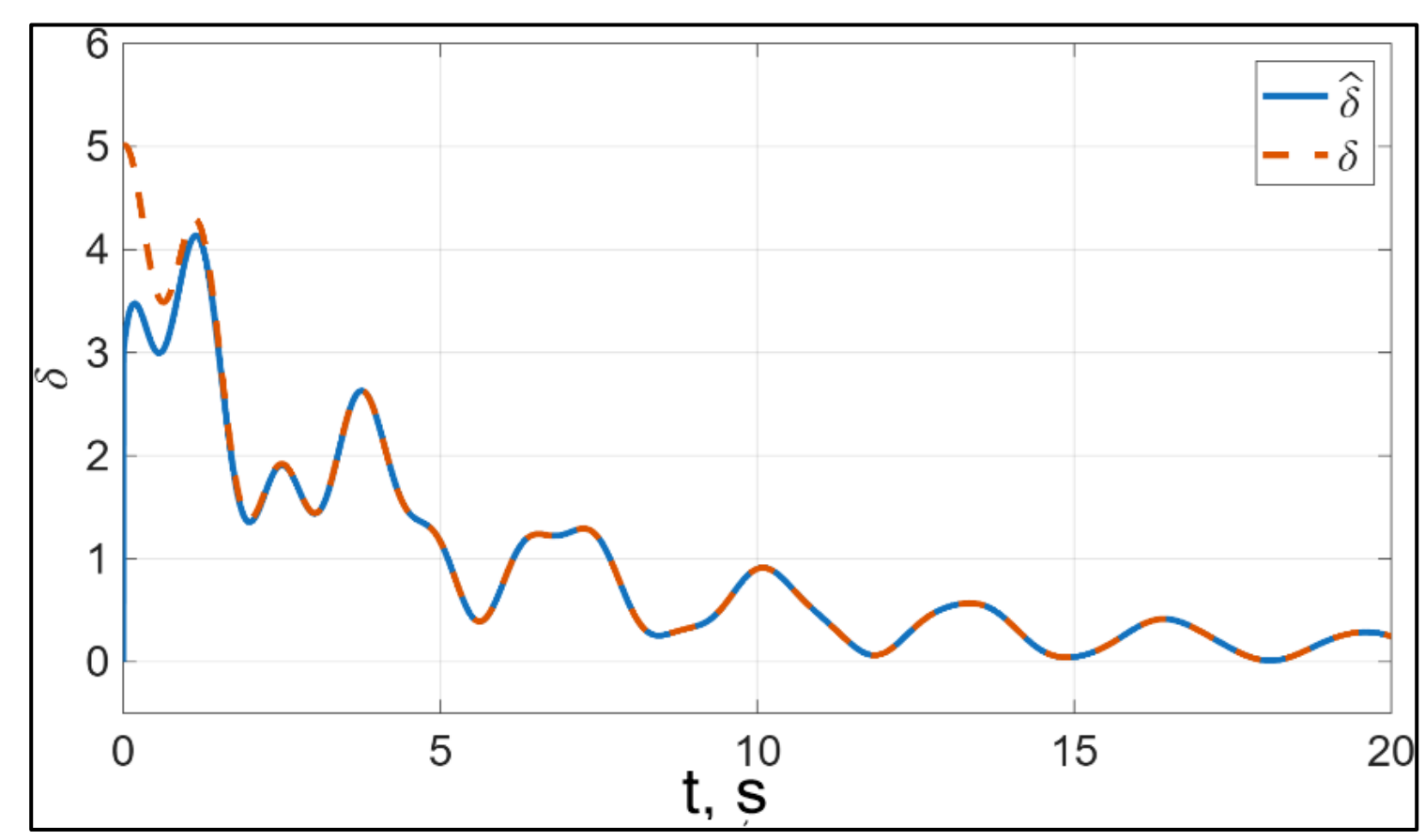

Figure 2 – Estimation results for $\hat{\delta}$

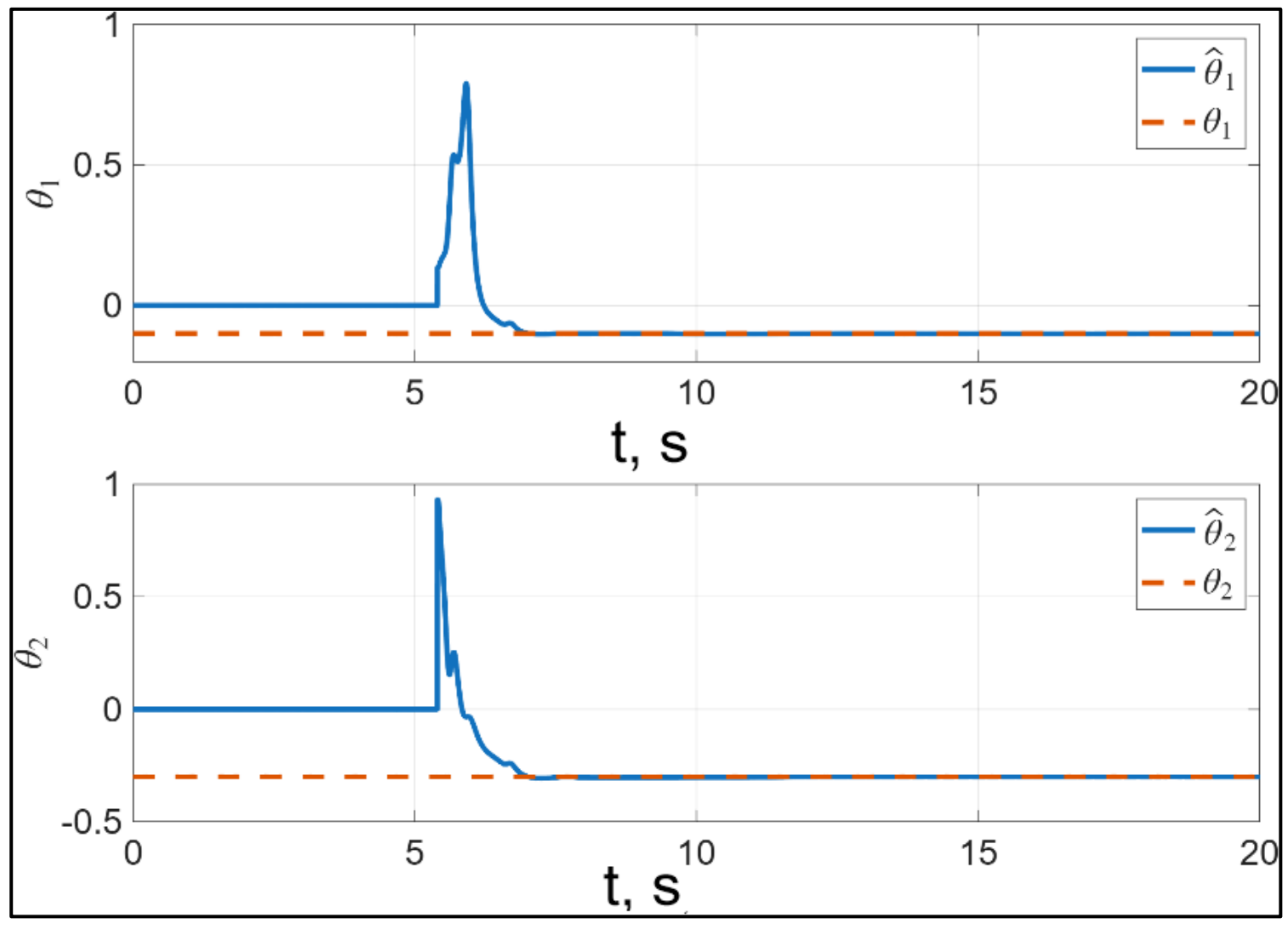

Figure 3 – Estimation results for $\hat{\theta}_1$ and $\hat{\theta}_2$

The simulation results demonstrate the effectiveness of the proposed algorithms under the introduced assumptions.

**Conclusion.** This paper presents a detailed observer design for a class of descriptor systems with unknown disturbances. It shows how to obtain a reduced dynamic equation, derive the corresponding reduced output equation, and formulate structural conditions that eliminate the disturbance from the observer for $x_a$. After the estimate $\hat{\boldsymbol{x}}_a$ is obtained, the disturbance signal is reconstructed, and a procedure for estimating its unknown parameters is demonstrated. The proposed parameterization method enables the nonlinear dependence on unknown parameters to be transformed into a linear regression with respect to certain combinations of these parameters. Numerical simulation confirms convergence of the estimation error for the dynamic component of the state vector to zero, whereas the estimation errors for the disturbance and its parameters converge to a neighborhood of zero. The obtained disturbance parameters estimates can be used to construct a stabilizing controller based on the internal model principle.

## REFERENCES

1. Duan, Guang-Ren. Analysis and design of descriptor linear systems. Vol. 23. Springer Science & Business Media, 2010.

2. Kunkel, Volker Mehrmann Peter, and Volker Mehrmann. Differential-algebraic equations. Zürich: European Mathematical Society, 2006.

3. Lamour, René, Roswitha März, and Caren Tischendorf. Differential-algebraic equations: a projector-based analysis. Springer Science & Business Media, 2013.

4. Gerdin, Markus. Parameter estimation in linear descriptor systems. Univ., 2004.

5. A. Ilchmann and T. Reis, Surveys in Differential-Algebraic Equations I. Springer Berlin, 2015

6. Campbell, Stephen, et al., eds. Applications of differential-algebraic equations: examples and benchmarks. Cham: Springer International Publishing, 2019.

7. Biehn, N., et al. "Observer design for linear time varying descriptor systems: numerical algorithms." Proceedings of the 37th IEEE Conference on Decision and Control (Cat. No. 98CH36171). Vol. 4. IEEE, 1998.
8. Konovalov, D., et al. "Finite-time observer design for linear descriptor systems." 2023 31st Mediterranean Conference on Control and Automation (MED). IEEE, 2023.
9. Moysis, Lazaros, et al. "Observer design for rectangular descriptor systems with incremental quadratic constraints and nonlinear outputs—Application to secure communications." International Journal of Robust and Nonlinear Control 30.18 (2020): 8139-8158.
10. Tripathi, Meenakshi, et al. "Observer design for nonlinear descriptor systems: A survey on system nonlinearities." Circuits, Systems, and Signal Processing 43.5 (2024): 2853-2872.
11. Ortega, Romeo, et al. "Adaptive state observers of linear time-varying descriptor systems: A Parameter estimation-based approach." IEEE Transactions on Automatic Control (2025).
12. Nugroho, Sebastian A., et al. "Observers for differential algebraic equation models of power networks: Jointly estimating dynamic and algebraic states." IEEE transactions on control of network systems 9.3 (2022): 1531-1543.
13. P. Colaneri, G. P. Incremona and L. Mirkin, "On Internal Model Compensation for General Regulator Problems," 2023 62nd IEEE Conference on Decision and Control (CDC), Singapore, Singapore, 2023, pp. 2657-2662.
14. P. Colaneri, G. P. Incremona and L. Mirkin, "A Descriptor Approach to Compensating Internal Models," 2024 IEEE 63rd Conference on Decision and Control (CDC), Milan, Italy, 2024, pp. 2916-2921
15. Bobtsov, A. A., Oskina, O. V., Slita, O. V., Nikolaev, N. A., & Boitsev, A. A. (2026). Harmonic carrier frequency estimation of a disturbed amplitude modulated signal. Scientific and Technical Journal of Information Technologies, Mechanics and Optics, 26(1), 214-217.

16. Margun, A. A., Bui, V. H., Bobtsov, A. A., & Efimov, D. V. (2025). State estimation for a class of nonlinear time-varying uncertain systems under multiharmonic disturbance. European Journal of Control, 82, 101185.

17. Aranovskiy, S., Bobtsov, A., Ortega, R. and Pyrkin, A., 2016. Performance enhancement of parameter estimators via dynamic regressor extension and mixing. IEEE Transactions on Automatic Control, 62(8), pp.3546-3550.

18. Sastry S., Bodson M. Adaptive control: stability, convergence and robustness. – Courier Corporation, 2011.